\documentclass[pra,aps,twocolumn,superscriptaddress,nofootinbib]{revtex4}
\usepackage{bm,graphicx,amsmath}
\usepackage{bbm}

\usepackage{amssymb}
\usepackage{epstopdf}
\usepackage{amstext}
\usepackage{latexsym}

\newcommand{\bk}[2]{\langle#1|#2\rangle}

\newcommand{\ket}[1]{\left\vert#1\right\rangle}
\newcommand{\bra}[1]{\left\langle#1\right\vert}

\begin{document}

\title{Ramsey interferometry with a spin embedded in a Coulomb chain}
\author{Gabriele De Chiara}
\affiliation{Grup d'{\`O}ptica, Departament de F{\'i}sica, Universitat Aut{\`o}noma
de Barcelona, E-08193 Bellaterra, Spain}
\author{Tommaso Calarco}
\affiliation{Institut f\"ur Quanteninformationsverarbeitung, Universit\"at Ulm, D-89069 Ulm, Germany}
\affiliation{European Centre for Theoretical Studies in Nuclear Physics and Related Areas, I-38050 Villazzano (TN), Italy}
\author{Shmuel Fishman}
\affiliation{Department of Physics, Technion, 32000 Haifa, Israel}
\author{Giovanna Morigi}
\affiliation{Grup d'{\`O}ptica, Departament de F{\'i}sica, Universitat Aut{\`o}noma
de Barcelona, E-08193 Bellaterra, Spain}

\date{\today}

\begin{abstract} We show that the statistical properties of a
Coulomb crystal can be measured by means of a standard
interferometric procedure performed on the spin of one ion in the
chain. The ion spin, constituted by two internal levels of the
ion, couples to the crystal modes via spatial displacement induced
by photon absorption. The loss of contrast in the interferometric
signal allows one to measure the autocorrelation function of the
crystal observables. Close to the critical point, where the chain
undergoes a second-order phase transition to a zigzag structure,
the signal gives the behaviour of the correlation function at the
critical point.\end{abstract} \maketitle

\section{Introduction}

The quest for control of quantum dynamics of systems with
increasing size is one of the present challenges in technological
applications of quantum mechanics~\cite{PZroadmap}. It involves
understanding the transition from the quantum to the classical
world~\cite{zurek-today} and is based on the full knowledge of how
the quantum properties scale with the system size, and in
particular, of how the thermodynamic properties are related to the
system microscopic dynamics~\cite{FordKacMazur,leggett-rmp,weiss}.
Several experiments pursue a bottom-up approach, where systems of
increasing complexity are built by combining together simpler
systems, over which one has full control~\cite{PZroadmap}. In this
context, Coulomb crystals of ions in Paul and Penning traps
constitute a prominent system. These crystals are composed by cold
ions in a confining potential that balances the Coulomb repulsion.
The ions vibrate around fixed positions in analogy to the
situation in an ordinary solid, while the interparticle distance
is usually of the order of several micrometers, constituting an
extremely rarefied type of condensed matter~\cite{DubinRMP}.
Varying the potential permits one to control the number of ions,
allowing one to explore structures of very different sizes, thus
offering the opportunity of studying the dynamics from few
particles to mesoscopic systems. Besides, these structures provide
promising applications, among others, for quantum information
processors~\cite{QIPC:Ions,QIPC:Ions:Exp} and
simulators~\cite{Wunderlich,Porras,porras-spin-boson,Schatz}.

\begin{figure}[!t] \includegraphics[scale=0.7]{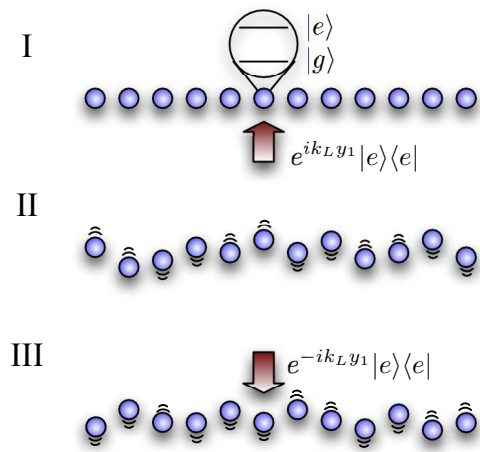}
\caption{\label{fig:1} Color online. Ramsey interferometry with a
chain. I. A transverse laser pulse prepares one ion of the chain
in a superposition of the internal states $|g\rangle$ and
$|e\rangle$. The mechanical effect, associated with the absorption
of a laser photon and conditioned to the ion being in state
$|e\rangle$, displaces the ion transversally and excites the modes
of the chain. II. The chain is let freely evolve for a time $t$.
III. A second laser pulse, addressing the same ion of the chain,
is sent in the opposite direction of the first pulse. The
probability that the ion is in state $|g\rangle$ is then measured
as a function of $t$.} \end{figure}

Among various crystalline structures experimentally
realized~\cite{Drewsen,Bollinger,Werth}, low-dimensional
structures, and in particular linear chains of ions, are routinely
produced in ion trap laboratories. Linear chains are obtained in
anisotropic potentials with large transverse confinement, usually
linear Paul traps~\cite{Ghosh}, and can be composed by several
dozens of ions. They exhibit a mechanical instability to a zigzag
structure as a function of the ion number or of the transverse
confinement, which has been first characterized experimentally
in~\cite{MPQ,Raizen}. The dynamics and thermodynamics of the
linear chain have been extensively studied in several theoretical
works~\cite{DubinPRE,Piacente,morigi2004}. Most recently, it has
been demonstrated that in the classical regime the mechanical
instability of the linear chain, which leads to the abrupt
transition to a zigzag structure, is a second order phase
transition~\cite{Dubin93,fishman2008}. The effects of quantum
fluctuations on low-dimensional Wigner crystals have been
discussed in~\cite{Schulz}, and are expected to be in general
negligible for Coulomb crystals of atomic ions~\cite{Yin}. Most
recently, in~\cite{retzker} parameter regimes have been discussed,
where quantum effects in structures of few ions could be
experimentally observed. These theoretical predictions lead to a
practical issue, namely how to determine experimentally the
thermodynamic quantities of large crystals. In fact, while the
quantum state of few trapped ions is experimentally fully
accessible by quantum tomographic techniques~\cite{Leibfried96, Measure:Ion},
the application of such technologies to larger crystals requires
diverging experimental times.

In this paper, we discuss a method for measuring the dynamical
and statistical properties of an ion chain, which is sketched in
Fig.~\ref{fig:1}. This method is based on an extension of Ramsey
interferometry with ions~\cite{Ramsey,Poyatos95}, which is applied
to a dipolar transition of one ion belonging to the chain. This
transition couples to the crystal modes via spatial displacement
induced by photon absorption or emission. By measuring the
interferometric contrast in a suitable setup, we show that one can
directly measure the autocorrelation function of the crystal. In
particular, we determine the dependence of the interferometric
signal on the system size and on the distance of the control
parameters from the critical point of the phase transition to the
zigzag configuration. Based on the latter dependence, the
interferometric signal may allow one to measure the critical
exponents.

This work is organized as follows. The salient properties of
Ramsey interferometry with ions in traps are reviewed in
Sec.~\ref{Sec:Interferometry}. In Sec.~\ref{Sec:Bath} we introduce
the system, composed by a Coulomb chain and the spin of one of the
ions of the crystal. The interferometric signals obtained for this
system are presented in Sec.~\ref{Sec:Signal}. The conclusions are
discussed in Sec.~\ref{Sec:Conclusions}. In the Appendix we report
the details of calculations at the basis of the results presented
in Sec.~\ref{Sec:Signal}.

\section{Ramsey Interferometry with ions}
\label{Sec:Interferometry}

Ramsey interferometry is routinely applied in quantum optics,
nuclear magnetic resonance, atomic and molecular physics for high
precision measurements~\cite{Ramsey}. Most recently, it has been
considered for measuring the loss of coherence in quantum optical
systems due to coupling with an external environment. Prominent
experimental applications are in cavity quantum
electrodynamics~\cite{Ramsey:CQED} and trapped
ions~\cite{Sackett}. The idea at the basis of this implementation
can be summarized as follows. Loss of coherence in quantum
dynamics is often associated to the partition of a large quantum
system into subsystems, of which only one part $S$ is fully
controllable in terms of unitary operations and/or of measurements
performed, while the rest, $R$, acts as a reservoir. The system
$S$ can be for instance a two-level transition or a spin, on which
interferometry is applied. In suitably designed setups, one can
have spin-dependent coupling to the reservoir $R$, such that the
dynamics establish correlations between $S$ and $R$. Such
correlations give rise to effects mimicking noise and loss of
coherence in the dynamics of the observables for the spin degrees
of freedom~\cite{Paz:LesHouches}. Loss of coherence can be
revealed in the off-diagonal elements of the spin density matrix
as a function of time, which can be measured in the
interferometric signal. Their time evolution is directly related
to the creation of correlations with the reservoir $R$, and thus
also indirectly related to the statistical properties of the reservoir
itself. The behavior of the loss of contrast in the
interferometric signal, hence, also permits one to measure some
statistical properties of the reservoir. This concept has been
applied for determining quantum properties of cold atomic
gases~\cite{Demler}.

In this section we review the idea at the basis of the setup for
ion interferometry, as it was first introduced
in~\cite{Poyatos95}, and discuss the information the corresponding
interferometric signal may provide.

\subsection{Ion interferometry}
\label{Sec:Ion:Ramsey}

We consider a single atom of mass $m$ in a harmonic trap, whose
electronic states $|g\rangle$ and $|e\rangle$ (both metastable)
are resonantly coupled by lasers The two states $\ket g$ and $\ket
e$ can be, for example, hyperfine states of the optical quadrupole
transition S$_{1/2}\to$ D$_{5/2}$ in ${}^{40}$Ca$^+$ whose
lifetime is of the order of $1$ s~\cite{IonTrapReview}.
Alternatively one can couple two metastable states of a hyperfine
multiplet via a coherent Raman transition using two lasers, as in
Ref.~\cite{Leibfried96}, or use two states of a magnetic dipole
r.f.-transition as proposed in~\cite{Wunderlich}. Assuming that
only one direction of the motion is relevant to the dynamics, the
Hamiltonian of the system has the form \begin{equation} \label{H}
H=H_{\rm at}+H_{\rm B}+H_{\rm int}\,, \end{equation} where
\begin{equation} \label{eq:Hat} H_{\rm
at}=\hbar\varpi_e|e\rangle\langle e| \end{equation} is the
Hamiltonian for the internal transition at frequency $\varpi_e$,
{\it i.e.}, the system $S$, and $H_{\rm B}=\hbar \nu b^{\dagger}b$
is the Hamiltonian for the oscillator of the atom center of mass
motion along the $y$-direction, {\it i.e.}, the reservoir $R$ --
with $b$ and $b^{\dagger}$ the annihilation and creation
operators, respectively, of one excitation at energy $\hbar\nu$.
The term $H_{\rm int}$ describes the interaction of the two-level
system with a laser pulse at frequency $\omega_L$ and wave vector
$k_L$ along the $y$-direction, and reads \begin{equation} H_{\rm
int}=\hbar \Omega(t)\left[\sigma^{\dagger} {\rm e}^{-{\rm i}
(\omega_L t-k_L y)} + {\rm H.c.}\right]\,, \label{H:int}
\end{equation} where $\sigma=\ket g \bra e$, $\sigma^{\dagger}$
its adjoint, and $\Omega(t)$ is the real-valued Rabi frequency.
From now on we take $\omega_L=\omega_0$ and consider the dynamics
in the reference frame rotating at the laser frequency. In this
frame, the phase of the field depends solely on the center-of-mass
position $y$ of the ion. By absorbing/emitting a photon, the atom
experiences a mechanical displacement due to the field phase
gradient over the center-of-mass wave packet. This mechanical
action is described in Eq.~(\ref{H:int}) by the operator
$\exp({\rm i}k_Ly)$, which, for the harmonic motion, is the
displacement operator \begin{equation} \label{eq:displacement}
D(\alpha)=\exp(\alpha b^\dagger-\alpha^*b)\,,\end{equation} where
we used $y=a_0(b^{\dagger}+b)$ and $\alpha={\rm i}k_La_0$ with
$a_0=\sqrt{\hbar/2m\nu}$.

We now assume that at time $t=0$ the ion is prepared in the
internal and oscillator ground state
$|\Psi(0)\rangle=|g,0\rangle$. In the time interval $[0,t_{\rm
pulse}]$ a square pulse at constant intensity $\Omega(t)=\Omega_0$
is applied, such that $\Omega_0 t_{\rm pulse}=\pi/4$ (thereby
implementing a so-called $\pi/2$ pulse). In the regime in which
the pulse intensity is sufficiently large, such that $\nu t_{\rm
pulse}\ll 1$, the free evolution of the oscillator can be
neglected during the pulse and the ion is prepared in the
superposition \begin{equation} \label{Psi:T} |\Psi(t_{\rm
pulse})\rangle=\frac{1}{\sqrt{2}}\left(|g,0\rangle+|e,\alpha\rangle\right)\,,
\end{equation} where $\ket\alpha=D(\alpha)\ket 0$ is a coherent
state.

The system is then let freely evolve for a time $t$
while a unitary operation is applied, which introduces a
spin-dependent phase $\phi$ according to the procedure
\begin{eqnarray}
\ket g&\to& \ket g\,,\\
\ket e&\to& e^{i\phi}\ket e\,, \end{eqnarray} and that the
interferometric procedure aims at detecting. At time $t+t_{\rm
pulse}$, assuming a purely Hamiltonian dynamics, the state of the
system reads \begin{equation} |\Psi(t+t_{\rm
pulse})\rangle=\frac{1}{\sqrt{2}}\left(|g,0\rangle+{\rm e}^{{\rm
i}\phi}|e,\alpha {\rm e}^{-{\rm i}\nu
t}\rangle\right)\,.\label{Psi:t} \end{equation} A squared laser
pulse is then switched on for a duration $3 t_{\rm pulse}$,
thereby implementing a $-\pi/2$ pulse on the spin.

In the absence of coupling with the motional degrees of freedom,
the probability to find the atom in the initial state after this
pulse would solely depend on the phase $\phi$. However, due to the
coupling with the external degrees of freedom, there is a
dephasing which arises from the free evolution of the coherent
state $\ket\alpha$. As a result, the probability to measure the
atom in the ground state after the last pulse takes the explicit
form \begin{eqnarray} \mathcal P_g(t)=\frac{1}{2}\left[1+{\rm
Re}\left\{ {\rm e}^{{\rm i}\phi}\mathcal S (t)\right\}\right]
\label{eq:prg:1}\,, \end{eqnarray} with \begin{eqnarray} \mathcal
S (t)&=&
\langle \alpha|\alpha{\rm e}^{-{\rm i}\nu t}\rangle\nonumber\\
&=&{\rm e}^{{\rm i}|\alpha|^2 \sin\nu t} {\rm
e}^{-2|\alpha|^2\sin^2(\nu t/2)} \end{eqnarray} and $0\le
|\mathcal S (t)|\le 1$. For fixed time evolution $t$, the
probability $\mathcal P_g$ as a function of $\phi$ is hence a
periodic signal exhibiting fringes with visibility
\begin{equation} \label{eq:visibility} \mathcal V=|\mathcal S
(t)|. \end{equation} In~\cite{Sackett} a version of this
interferometer has been used in order to measure decoherence of
the ion quantum motion, due to the coupling to a classical noise
applied at the electrodes of a Paul trap. The interferometric
signal hence gave a measure of how coherence decays as a function
of time for different statistical properties of the applied noise.

\subsection{Interferometric signal and reservoir properties}
\label{Sec:Englert}

Let us now discuss the physical meaning of the quantity $\mathcal
S(t)$ in Eq.~(\ref{eq:prg:1}). From an interferometric point of
view, we notice that loss of visibility is found at times $t$ that
are different from an integer multiple of the oscillation period
of the ion. At these instants, then, the $-\pi/2$ pulse does not
bring the spin back to the initial state, and there are residual
correlations between ion motion and spin which gives rise to a
diminution of the contrast. In other words, a which-way
information is left in the reservoir $R$ after it interacted with
the system. This connection has been elegantly elucidated by
Englert in~\cite{Englert}, where a physical quantity, the
distinguishability $\mathcal D$, was introduced in order to
quantify the amount of which-way information left by means of a
unitary evolution coupling spin and reservoir. Englert showed that
visibility and distinguishability are related by the
inequality~\cite{Englert} \begin{equation} {\mathcal
D^2}+{\mathcal V^2}\le 1\,, \end{equation} where the equality
holds when the reservoir is initially in a pure state.

We also notice that the visibility of the signal in
Eq.~(\ref{eq:visibility}) can be rewritten as  \begin{equation}
\mathcal V(t) =\left|\bra 0 e^{-ik_Ly(t)} e^{ik_Ly(0)}  \ket 0
\right|= \exp\left[-\frac{k_L^2}{2} \mathcal G(t)\right]
\,,\end{equation} where \begin{equation} \label{eq:G} \mathcal
G(t)=\langle [y(t)-y(0)]^2\rangle \end{equation} is the
autocorrelation function of the reservoir, and the mean value
$\langle .\rangle$ is taken over the initial state of $R$. The
visibility is hence a direct measure of the autocorrelation
function of the reservoir $R$. This result holds in general, when
the state of the reservoir $R$ is Gaussian. Designing a different
setup will give access to different correlation functions of the
reservoir. By implementing a mechanical excitation as
in~\cite{Wunderlich}, for instance, one can measure the
anti-symmetric part of the correlation function. Alternatively, by
applying the pulses to two different ions one can determine the
correlation length of the crystal. In general, the Ramsey signal
will allow one to characterize the thermodynamic properties of a
reservoir $R$ in thermal equilibrium.

\subsection{Discussion} \label{sec:discussionspin} The visibility
of the interference signal is reminiscent of the Loschmidt echo as
defined in \cite{cucchietti03}. There, a quantity is evaluated
which is the overlap between two states evolved from the same
wavefunction under the influence of two different Hamiltonians,
and which is analogous to $\mathcal S(t)$ in Eq.~\eqref{eq:prg:1}.
Such an overlap is directly related to the Loschmidt echo, which is
frequently used to describe loss of coherence as a consequence of
the interaction between a system and its environment. The
Loschmidt echo has been extensively analyzed for one-dimensional
spin chains, where the system is one spin and the decohering
environment is the rest of the chain, constituting a spin bath and
characterized by Ising or Heisenberg nearest-neighbor
interactions~\cite{spin-rossini}. Differing from that model, the
present work is concerned with an environment characterized by
long-range interactions, which emerge from the Coulomb repulsion
between the ions of the chain. However, we shall show that the
Loschmidt echo in~\cite{spin-rossini} and the visibility of the
Ramsey signal discussed in this work exhibit important
similarities when the corresponding environment is close to a
critical point, marking a second-order phase transition.

\section{A spin coupled to a Coulomb chain} \label{Sec:Bath}

The theoretical model, which we analyze in this work, is a Coulomb
chain coupled to the two-level electronic transition of one of the
ions composing the chain, which we will denote as the spin.
Interaction between the spin and the chain occurs via the
mechanical effects of light: The ion displacement, due to photon
absorption or emission, induces a mechanical excitation of the
chain modes via the Coulomb and trap forces binding the crystal.
The Hamiltonian of the system reads \begin{eqnarray} H=H_{\rm
at}+H_{\rm B}^{(N)}+H_{\rm int}\,, \end{eqnarray} where $H_{\rm
at}$ is the spin Hamiltonian given in Eq.~\eqref{eq:Hat}, $H_{\rm
B}^{(N)}$ describes the dynamics of the chain, and $H_{\rm int}$
gives the interaction of the spin with the laser pulse and is
defined in Eq.~\eqref{H:int}. In the first part of this section we
review the basic properties of the ion chain. In the second part
we characterize the mechanical effects of light on the crystal due
to the coupling with the laser pulse, described by the term
$H_{\rm int}$.

\subsection{The Coulomb chain}

The Coulomb chain is composed by $N$ ions, each having the same
mass $m$ and charge $Q$, and confined inside a highly anisotropic
trap, which stabilizes the crystalline order along a
line\footnote{Experimentally, ion chains are usually realized in
linear Paul traps. Some experiments reported ion chains in ring
traps, as for instance in~\protect\cite{MPQ}. Here,
crystallization is achieved combining laser cooling, to the regime
in which the ions thermal energy is much smaller than the
interaction energy, with some sort of pinning.}. For convenience,
in the rest of this paper we will consider a ring trap of very
large radius. In this regime the system is equivalently described
by a linear distribution of charges on a line of length $L_{\rm
chain}$ with periodic boundary conditions, where the ions
experience a mutual repulsive interaction along the linear axis,
which we identify with the $x$-axis, while the transverse
direction is confined by a steep harmonic potential of frequency
$\nu_t$~\cite{JPB}. We assume that one ion is pinned at the
origin, so that the equilibrium positions of the other ions are
${\bf r_j^{(0)}}=(x_j^{(0)},0,0)$ with $x_j^{(0)}=ja$, $j=0,
1,\ldots, N$, and $a$ is the interparticle distance, $a=L_{\rm
chain}/N$. Moreover, we assume that the ions are sufficiently cold
such that their vibrations around the equilibrium points are
harmonic. We denote by $q_j=x_j-x_j^{(0)}$, $y_j$, and $z_j$ the
axial and transverse displacements of the ion $j$ from its
equilibrium position. Their dynamics are then described by the
Hamiltonian $$H_{\rm B}^{(N)}=V_0+H_x+H_y+H_z,$$ where $V_0$
denotes the energy of the classical ground state and $H_{\ell}$
($\ell=x,y,z$) give the harmonic motion about the classical
equilibrium positions. In the normal modes decomposition, after
quantizing the linear excitations, these terms take the
form~\cite{morigi2004} \begin{eqnarray}
&&H_\ell=\sum_{k}\sum_{\sigma=\pm}
\hbar\omega_{\ell}(k)\left(b_{\ell}^{\dagger}(k,\sigma)b_{\ell}(k,\sigma)+\frac{1}{2}\right)\,,
\label{H:y} \end{eqnarray} where $k$ labels the mode wave vector,
which takes the values $k=2\pi n/Na$ with $n=0,\ldots,N/2$, while
$\sigma=\pm$ denotes the mode parity (see also Appendix A). The
operators $b_{\ell}(k,\sigma)$ and $b_{\ell}^{\dagger}(k,\sigma)$
annihilate and create, respectively, a quantum of energy
$\hbar\omega_{\ell}(k)$ of the corresponding oscillator,
whereby~\cite{Ashcroft,fishman2008} \begin{eqnarray}
&&\omega_{x}(k)^2=
8\omega_0^2\sum_{j=1}^{N/2}\frac{1}{j^3}\sin^2\frac{jka}{2}\;,
\label{eq:linspectrum1}\\
&&\omega_{y}(k)^2=\omega_{z}(k)^2=
\nu_t^2-4\omega_0^2\sum_{j=1}^{N/2}\frac{1}{j^3}\sin^2\frac{jka}{2}\;,
\label{eq:linspectrum2}
 \end{eqnarray} with
\begin{equation} \label{omega:0} \omega_0^2 = \frac{Q^2}{ma^3}
\end{equation} a characteristic frequency scaling the spectral
frequencies of the crystal. We notice that the transverse trap
frequency $\nu_t$ is the largest eigenfrequency of the transverse
modes.

The displacements $q_j,y_j,z_j$ of the ion at position ${\bf
r_j^{(0)}}$ are related to the normal modes by the orthogonal
matrix ${\cal R}$, such that~\cite{Morigi-Walther,Morigi-Doppler}
\begin{eqnarray}
&&q_j=\sum_{k,\sigma}\sqrt{\frac{\hbar}{2m\omega_x(k)}} {\cal
R}_{j,k\sigma}\left[b_x(k,\sigma)+b_x^{\dagger}(k,\sigma)\right]\,,
\\
&&y_j=\sum_{k,\sigma} \sqrt{\frac{\hbar}{2m\omega_y(k)}}{\cal
R}_{j,k\sigma}\left[b_y(k,\sigma)+b_y^{\dagger}(k,\sigma)\right]\,,
\label{y:j} \end{eqnarray} and analogously for $z_j$. The specific
form of the orthogonal matrix ${\cal R}$ is given in
Appendix~\ref{App:A}.

The mechanical stability of the ion chain is warranted provided
that the transverse confinement satisfies the relation
$\nu_t>\nu_t^{(c)},$ where $\nu_t^{(c)}$ is a critical frequency
which depends on $\omega_0$ and on the density of ions $1/a$. For
ions on a ring one finds~\cite{Dubin93,fishman2008}
\begin{equation} \nu_t^{(c)}
=\omega_0\sqrt{\frac{7}{2}\zeta(3)}\,. \end{equation} At
$\nu_t=\nu_t^{(c)}$ the chain undergoes a second-order phase
transition to a planar structure, such that for
$\nu_t<\nu_t^{(c)}$ the ions organize in a zigzag structure across
the $x$-axis. Assuming that the zigzag structure is localized on
the $x-y$ plane, the new equilibrium positions are ${\bf
r_j^{(0)}}=(x_j^{(0)},y^{(0)}_j,0)$ where $x_j^{(0)}=ja$,
$y^{(0)}_j = (-1)^j b/2$, and $z^{(0)}_j =0$. The transverse
displacement $b$ depends on the linear density of ions $a$ and on
the value of $\nu_t$, as shown in Ref.~\cite{fishman2008}. In the
zigzag configuration, the Hamiltonian describing the harmonic
oscillations around the equilibrium positions takes the form
$$H_{\rm B}^{(N)}=V_0^{\rm zz}+H_{xy}^{\rm zz}+H_z^{\rm zz}\,,$$
where $V_0^{\rm zz}$ is the energy of the new classical ground
state. The Hamiltonian $H_{xy}^{\rm zz}$ is the harmonic term for
the displacements $q_j$ and $w_j=y_j-y_j^{(0)}$ around the new
equilibrium positions, which are now coupled. The dispersion
relation is now defined in the Brillouin zone $[0,\pi/2a]$, such
that the wave vector $k$ takes the values $k=2\pi n/Na$ with
$n=0,\ldots,N/4$. The spectrum  of the excitations along $x$ and
$y$ presents four branches, which we label by $\beta=1,2,3,4$ (a
detailed discussion of the form of the spectrum can be found
in~\cite{fishman2008}). In the normal modes basis Hamiltonian
$H_{xy}^{\rm zz}$ reads \begin{eqnarray} \label{eq:hamcoordinates}
H_{xy}^{\rm zz}=\sum_{k}\sum_{\sigma=\pm}\sum_{\beta=1}^4\hbar
\omega_{\beta}^{\rm
zz}(k)\left[c_{\beta}^{\dagger}(k,\sigma)c_{\beta}(k,\sigma)+\frac{1}{2}\right],
\end{eqnarray} where $\sigma$ is the mode parity. Finally, the
operators $c_{\beta}(k,\sigma)$ and
$c_{\beta}^{\dagger}(k,\sigma)$ annihilate and create,
respectively, an excitation of the mode of the branch $\beta$ with
wave vector $k$, frequency $\omega_{\beta}(k)$, and parity
$\sigma$. Hamiltonian $H_z^{\rm zz}$ is the harmonic term for the
displacement $z_j$ and is not explicitly reported, as it will not
be relevant for the rest of this paper.

The displacements $q_j$, $w_j$ are related to the normal modes by
the orthogonal matrix ${\cal R}^{\rm zz}$ according to the
relations \begin{equation} \label{eq:rho}
\varrho_j=\sum_{k}\sum_{\sigma=\pm}
\sum_{\beta}\sqrt{\frac{\hbar}{2m\omega_{\beta}^{\rm zz}(k)}}
{\cal R}^{\rm
zz}_{j,k\beta\sigma}\left[c_{\beta}(k,\sigma)+c_{\beta}^{\dagger}(k,\sigma)\right]\,,
\end{equation} where $\varrho=(q_1,w_1,\ldots,q_N,w_N)$ and the
index $j=1,\ldots,2N$. The specific form of the orthogonal matrix
for the zigzag depends on the value of $\nu_t$ and it is reported
in Appendix~\ref{App:A}.

In this work, we will also be interested in studying the
interferometric signal as a function of the distance of the
transverse frequency from the critical value $\nu_t^{(c)}$. We
will then denote this quantity by the parameter \begin{equation}
\label{Ew:Delta} \Delta=\nu_t-\nu_t^{(c)}, \end{equation} which
has the dimension of an angular frequency. In particular, when
$\Delta>0$ the linear chain is stable, while for $\Delta<0$ the
system is found in the zigzag
configuration\footnote{The structure is found in a zigzag
configuration for values of $\nu_t$, such that
$\nu_t^{(c)}>\nu_t>{\tilde\nu}_t^{(c)}$, where
${\tilde\nu}_t^{(c)}$ is another critical value at which the
system undergoes a phase transition to a multiple chain
structure~\cite{MPQ,Dubin93,Piacente}.}.

\subsection{Ramsey interferometry with the Coulomb chain}

Ramsey interferometry with the Coulomb chain is implemented by
means of a straightforward extension of the procedure described in
Sec.~\ref{Sec:Ion:Ramsey} for a single ion. The basic setup
discussed in this paper is sketched in Fig.~\ref{fig:1}. We
assume that the ion at position ${\bf r_1}$ in the chain is
selectively addressed by the laser pulses, which propagate
perpendicularly to the chain, and restrict the Hilbert space of
the electronic degrees of freedom to the two level of its internal
transition. By absorbing and emitting a photon during the pulse,
the ion undergoes a transverse displacement, which excites all
modes of the chain coupling with the ion position. As in
Sec.~\ref{Sec:Ion:Ramsey}, we assume that the duration $t_{\rm
pulse}$ of the laser pulses is sufficiently short, such that the
free evolution of the chain during the pulse is negligible. This
corresponds to imposing the relation $\omega_\textrm{max} t_{\rm
pulse}\ll 1$, where $\omega_\textrm{max}$ is the largest
eigenfrequency of the crystal modes that are involved in the
dynamics. In this paper we consider laser pulses propagating
along the $y$-direction, inducing hence mechanical displacement
along $y$. Hence, when the ions form a linear chain $\omega_{\rm
max}=\nu_t$ and the condition to be fulfilled is $\nu_tt_{\rm
pulse}\ll 1$.

The ion displacement due to absorption or emission of a photon is
described by the operator
\begin{eqnarray}
\exp(i k_L y_1) &=&
\prod_{k,\sigma} \exp\left[i \eta(k,\sigma)
(b_y(k,\sigma)+b_y^\dagger(k,\sigma)) \right]
\nonumber\\
&=& \prod_{k,\sigma} D_{k,\sigma}(\alpha_{k\sigma})\,,\label{D:N}
\end{eqnarray}
 which has been here conveniently expressed in terms
of the normal modes of the linear chain using Eq.~(\ref{y:j}).
Here \begin{equation}
\eta(k,\sigma)=k_L\sqrt{\frac{\hbar}{2m\omega_y(k)}}{\cal
R}_{1,k\sigma} \end{equation} is the Lamb-Dicke parameter for the
mode $\omega_y(k)$ scaling the displacement of the mode by photon
absorption or emission~\cite{Morigi-Doppler,Morigi-Walther}, and
$D_{k,\sigma}(\alpha_{k\sigma})$ denotes the displacement operator
for the oscillator at frequency $\omega_y(k)$ and parity $\sigma$
with amplitude \begin{eqnarray} \label{eq:alphaksigma}
\alpha_{k\sigma} = i\eta(k,\sigma) \,. \end{eqnarray} For later
convenience we denote by \begin{eqnarray}
|\{\alpha_{k,\sigma}\}\rangle=\prod_{k,\sigma}
D_{k,\sigma}(\alpha_{k\sigma})|0\rangle
&=&\bigotimes_{k,\sigma}\ket{\alpha_{k\sigma}(t)} \end{eqnarray}
the quantum state of the linear chain, obtained by applying the
displacement operator~(\ref{D:N}) to the ground state of the chain
$|0\rangle$.

Let us now consider some useful relations. The chain will be in
the Lamb-Dicke regime, {\it i.e.}, the ion displacement by photon
recoil perturbs weakly the mechanical state of the chain, provided
that the relation $|\eta(k,\sigma)|\sqrt{2n_{k\sigma}+1}\ll 1$ is
satisfied for all modes, where $n_{k\sigma}=\langle
b_y^{\dagger}(k,\sigma)b_y(k,\sigma)\rangle$ is the mean
occupation of the mode with frequency $\omega_y(k)$.
Correspondingly, $|\alpha_{k\sigma}|\ll 1$.

In general, in the rest of this work we will assume $\eta_y\ll 1$,
where $\eta_y=\eta_0/\sqrt{N}$ is the Lamb-Dicke parameter of the
transverse bulk mode at frequency $\nu_t$, and \begin{equation}
\eta_0=k_L\sqrt{\frac{\hbar}{2m\nu_t}}\,,\end{equation} while we
will not make any particular assumption for the Lamb-Dicke
parameter of the other modes. In particular, close to the
mechanical instability to the zigzag configuration, the Lamb-Dicke
parameter of the modes whose frequencies vanish, $\omega_y(k)\to
0$, can become very large. Hence, in such a regime, the photon
recoil can perturb significantly the state of the chain.

Finally, when studying the interferometric signal we will compare
curves obtained at different values of $\nu_t$. In this case we
will make use of the quantity \begin{equation} \label{eta:c}
\eta^{(c)}=k_L \sqrt{\frac{\hbar}{2m\nu_t^{(c)}}}\,,
\end{equation} which is related to the Lamb-Dicke parameter
$\eta_0$ by the equation
$\eta^{(c)}=\eta_0\sqrt{\nu_t/\nu_t^{(c)}}$. We conclude this
section by giving some typical values of the parameters that are
relevant in this work. In the experiment of Ref.~\cite{MPQ}, ion
chains of dozens of $^{24}$Mg$^+$ ions were realized, with
interparticle distance of the order of $a\sim 33\; {\mu}m$. This
case corresponds to the value of the typical frequency $\omega_0\sim
2\pi \times 64\;$ kHz, while the single particle Lamb-Dicke
parameter, close to the critical point, is $\eta_0\sim 0.62$.

\begin{figure}[!t] \begin{center}
\includegraphics[width=0.4\textwidth]{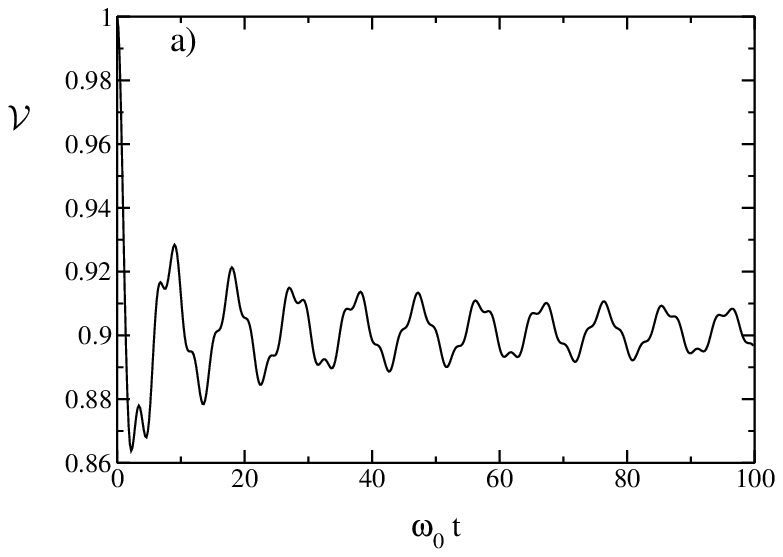}\\
\includegraphics[width=0.4\textwidth]{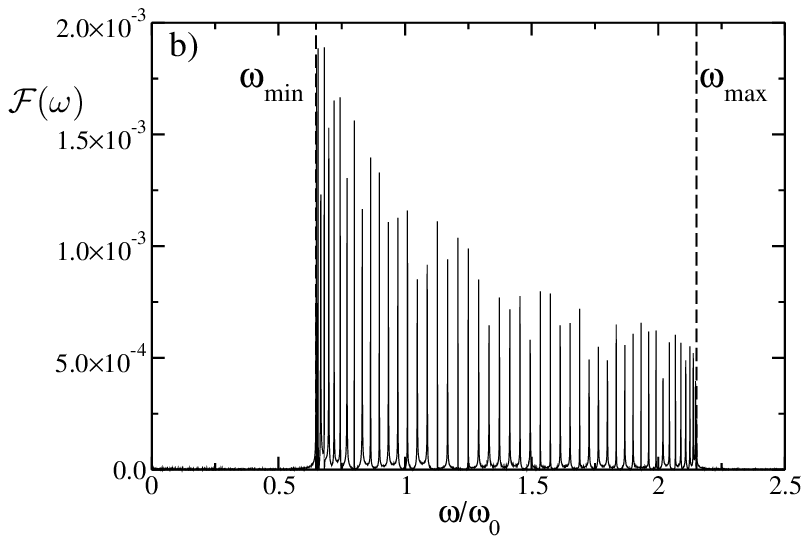} \caption{
\label{fig:prg1} a) Visibility $\mathcal V$ as a function of the
time elapsed between two Ramsey pulses, in units of $1/\omega_0$,
for $N=100$ ions in a linear chain. The parameters are $\nu_t =
\nu_t^{(c)}+\Delta$, with $\Delta=10^{-1}\omega_0$, and $
\eta^{(c)}= 0.25$. b) Fourier spectrum ${\mathcal F}(\omega)$ of
the visibility ${\mathcal V}(t)$, obtained by computing
numerically the discrete Fourier transform of ${\mathcal V}(t)$ in
a finite time interval $t\in[-T_F/2;T_F/2]$ with
$T_F=10^4/\omega_0$. The interval is sampled with a time slicing
$dt=T_F/n_s$ with $n_s=10^5$. We checked convergence of the
spectrum amplitude $\mathcal F$ by increasing $T_F$ and $n_s$. The
spectrum is normalized such that $\mathcal F(\omega=0)=1$. The
dashed lines indicate the minimum and maximum frequencies of the
transverse spectrum ($\omega_{\rm
min}=\sqrt{2\Delta\nu_t+\Delta^2}$ and $\omega_{\rm max}=\nu_t$).}
\end{center} \end{figure}

\begin{figure}[!t] \begin{center}
\includegraphics[width=0.4\textwidth]{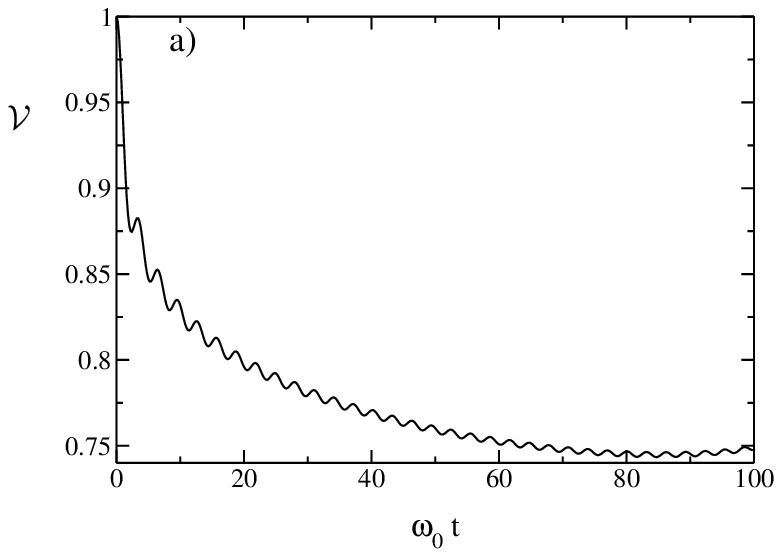}\\
\includegraphics[width=0.4\textwidth]{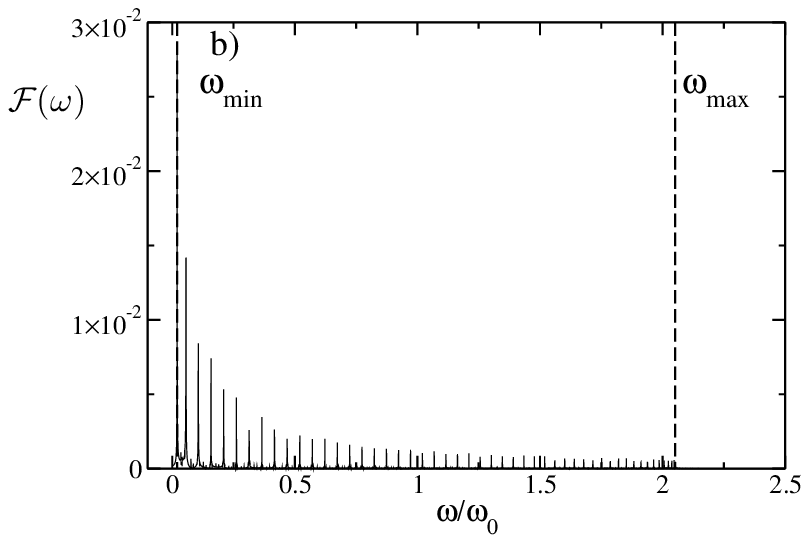} \caption{
\label{fig:prg2} Same as in Fig.~\protect\ref{fig:prg1} except for
$\nu_t=\nu_t^{(c)}+\Delta$ with $\Delta=10^{-4}\omega_0$. }
\end{center}\end{figure}

\section{Measuring the statistical properties of the crystal}
\label{Sec:Signal}

We now analyze the signal obtained by performing Ramsey
interferometry with the spin of the ion at the position ${\bf
r_1}$ of the chain. We assume that the spin is initially in the
electronic state $|g\rangle$ and that the chain has been prepared
in the ground state by some cooling
procedure~\cite{IonTrapReview}. Hence, the initial state in the
Hilbert space of the modes of the linear chain and of the spin
transition reads \begin{equation} \label{eq:psiin}
\ket{\Psi(0)}=\ket{g,0}\,. \end{equation} After the sequence of
Ramsey pulses, chain and spin are in the entangled state
\begin{equation} \label{Psi:Final}
|\Psi(t)\rangle=\ket{\varphi_g(t)}\ket g+i\ket{\varphi_e(t)}\ket
e\,, \end{equation} where $|\varphi_g\rangle$ and
$|\varphi_e\rangle$ are (not normalized) states of the crystal and
read \begin{eqnarray} \ket{\varphi_g(t)}&=& \frac{1}{2}\left[\ket
0+{\rm e}^{-{\rm i}ky_1}{\rm e}^{{\rm
i}\phi}\ket{\{\alpha_{k\sigma}(t)\}} \right]\,,
\\
\ket{\varphi_e(t)}&=&\frac{1}{2}\left[ \ket{\{\alpha_{k\sigma}\}}
-{\rm e}^{{\rm i}\phi}\ket{\{\alpha_{k\sigma}(t)\}}\right].
\end{eqnarray} Here,
$\alpha_{k\sigma}(t)=\alpha_{k\sigma}e^{-i\omega_y(k) t}$ and
$\phi$ is the phase due to the applied phase shift. The visibility
in the Ramsey signal is \begin{equation} \label{Eq:V} \mathcal
V=\left| \bk{ \{\alpha_{k\sigma}\} }{ \{\alpha_{k\sigma}(t)\} }
\right| =\exp[-A(t)]\,, \end{equation} where, for a linear chain,
the exponent reads \begin{equation} \label{eq:A}
A(t)=2\sum_{k\sigma} |\alpha_{k\sigma}|^2\sin^2\frac{\omega_y(k)
t}{2}\,. \end{equation} An analogous expression can be obtained
for the zigzag structure. The quantity $A(t)$ is directly related
to the autocorrelation function in Eq.~\eqref{eq:G},
\begin{equation} \label{eq:AG} A(t)=\frac{k_L^2}{2} \mathcal
G(t)\,, \end{equation} where this relation is valid both for the
linear chain and for the zigzag structure. Moreover, it also holds
when the crystal is in thermal equilibrium at temperature $T$. In
this case the function $A$ takes the form \begin{equation}
A_T(t)=2\sum_{k\sigma}\coth\frac{\hbar\omega_y(k)}{2k_B T}
|\alpha_{k\sigma}|^2\sin^2\frac{\omega_y(k) t}{2}\,,
\end{equation} where $k_B$ is the Boltzmann constant. In the rest of
the paper we will focus on the case $T=0$ for which the
second-order classical phase transition linear-zigzag at
$\nu_t=\nu_t^{(c)}$ occurs.

Figure~\ref{fig:prg1}a) displays the visibility $\mathcal V$ for
the linear chain as a function of the elapsed time $t$ for $N =
100$ ions and $\eta^{(c)}= 0.25$. The chain is sufficiently far
away from the critical point ($\Delta=10^{-1} \omega_0$), such
that all modes are in the Lamb-Dicke regime and the mechanical
effect of the photon perturbs weakly the chain stability.
Nevertheless, we observe that the visibility exhibits initially a
decay, reaching then values below unity, about which it oscillates
asynchronously for $t>0$. This behaviour is a clear consequence of
the time-dependent form of the signal, as it can be seen expanding
Eq.~(\ref{Eq:V}) in powers of the exponent $A(t)$, defined in
Eq.~(\ref{eq:A}): Hence, the Fourier spectrum of the visibility
contains in principle all possible sums of the eigenfrequencies
$\omega_y(k)$. The dispersion relation shows that the
eigenfrequencies are incommensurate, such that the signal in
Eq.~(\ref{eq:A}) is not periodic (although for finite systems it
may exhibit partial revivals in the excitations, as we will
discuss in Sec.~\ref{Sec:Asymptotics}). Following the
interferometric interpretation by Englert~\cite{Englert}, the
mechanical effect of light leaves a which-way information in the
chain excitations, which is only partially erased by the second
Ramsey pulse.

When the structure is in the Lamb-Dicke regime, the visibility
signal can be put in direct connection with the autocorrelation
function ${\mathcal G}(t)$, since ${\mathcal V}(t)\simeq 1-A(t)$.
The Fourier spectrum of the visibility, in this case, is thus just
composed by the frequencies of the normal modes.
Figure~\ref{fig:prg1}b) displays the Fourier spectrum of the
visibility in Fig.~\ref{fig:prg1}a), obtained by performing the
discrete Fourier transform of the signal ${\mathcal V}(t)$. The
sharp peaks, giving the main components of the Fourier spectrum,
correspond here to the frequencies $\omega_y(k)$ of the normal
modes.

Let us now discuss the signal when the parameters of the linear
chain are close to the critical point, where it undergoes a
transition to a zigzag configuration. Figure~\ref{fig:prg2}
displays the visibility and its Fourier spectrum  for
$\Delta=10^{-4} \omega_0$. The visibility signal decays to smaller
values as a function of the time elapsed between the two Ramsey
pulses, thereby exhibiting fast oscillations about the decaying
mean value. The Fourier spectrum shows that the low frequency
components of the transverse normal modes mostly contribute to the
signal. In particular, the lowest frequency mode, which
corresponds to the soft mode driving the instability to the
zigzag~\cite{fishman2008}, has the largest Fourier amplitude. In
fact, even if the chain is still in the Lamb-Dicke regime, the
Lamb-Dicke parameter of the soft mode is the largest, thus this
mode is mostly excited by the ion displacement due to photon
recoil. In this case, hence, the visibility allows one to access
the behaviour of the autocorrelation function close to the
critical point.

\subsection{Short times: quadratic decay of the visibility}
\begin{figure} \includegraphics[scale=0.25]{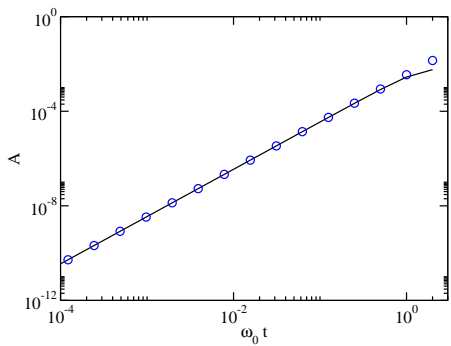}
\caption{\label{fig:shorttimes1} Color online. Function $A(t)$ as
a function of the time elapsed between two Ramsey pulses (in units
of $1/\omega_0$). The solid line corresponds to Eq.~\eqref{eq:A} and the symbols correspond to the short-time expansion in Eq.~\eqref{eq:Agauss}.
The curve has been evaluated for $\eta^{(c)}= 0.05$ in a linear
chain of $N=10^3$ ions. } \end{figure}

We now focus on the behaviour of the visibility for short elapsed
times $t$. In this regime the autocorrelation function ${\mathcal
G}(t)$, and correspondingly the logarithm of the visibility $A(t)$
in Eq.~\eqref{eq:A}, can be expanded in powers of $\omega_y(k)t\ll
1$. At lowest order in the expansion one finds \begin{equation}
\label{eq:Agauss} A\simeq\Gamma t^2\,, \end{equation} where
\begin{eqnarray} \Gamma &=& \frac12 \sum_{k,\sigma}
|\alpha_{k\sigma}|^2\omega_y(k)^2\,,
\label{eq:expansion-simple-ramsey} \end{eqnarray} and the
condition $\omega_{\rm max}t\ll 1$ must hold. The correlation
function hence goes quadratically with time, as shown in
Fig.~\ref{fig:shorttimes1}, and correspondingly the visibility
signal decays with a Gaussian-type behaviour. In particular, for
the linear chain the coefficient $\Gamma$ can be rewritten as
\begin{eqnarray} \label{Gamma} \Gamma= \frac{\hbar
k_L^2}{4m}\left( \frac{1}{N}\sum_{k,\sigma}\omega_y(k)\right)\,,
\end{eqnarray} and is hence proportional to the mean value of the
frequency of the transverse excitations. The detailed derivation
of this expression is reported in Appendix~\ref{App:B}.
Equation~(\ref{Gamma}) has been obtained for ions in a ring trap
with large radius. It approximates well the result for a linear
chain in a linear Paul trap when $N\gg 1$.
\begin{figure} \includegraphics[scale=0.65]{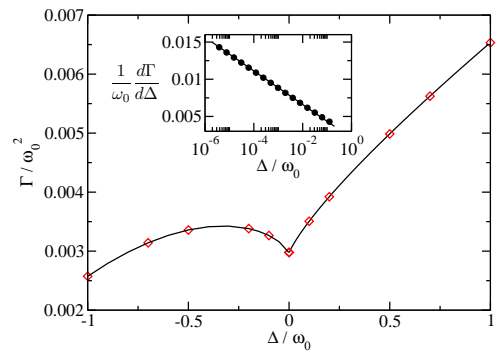}
\caption{\label{fig:shorttimes2} Color online. Coefficient
$\Gamma$ (in units of $\omega_0^2$) as a function of
$\Delta=\nu_t-\nu_t^{(c)}$ (in units of $\omega_0$).  The solid
line corresponds to the coefficient $\Gamma$, given in
Eq.~\eqref{eq:expansion-simple-ramsey} and the symbols correspond to
the value $\Gamma_{\rm fit}$, obtained by fitting the function
$A(t)$ in Eq.~(\ref{eq:A}) with the function $\Gamma_{\rm fit}
t^2$. The parameters are the same as in
Fig.~\ref{fig:shorttimes1}. Inset: Derivative of $\Gamma$ with
respect to $\Delta$ (in units of $\omega_0$) as a function of
$\Delta$ (in units of $\omega_0$ and logarithmic scale) for
$\Delta>0$ (symbols). An analytical analysis of the asymptotic behaviour, as
well as a numerical fit of the curve give a logarithmic dependence
of the derivative on $\Delta$ for $\Delta\to 0^+$ (line).}
 \end{figure}

We now analyze the dependence of the coefficient $\Gamma$ on
$\nu_t$ for values close to $\nu_t^{(c)}$,  thereby getting
further insight on the behaviour of the correlation function close
to the critical point. Figure~\ref{fig:shorttimes2} displays
$\Gamma$ as a function of $\Delta=\nu_t-\nu_t^{(c)}$ for values
close to the value $\Delta=0$. One clearly observes a minimum,
with a cusp-like behaviour, at the critical value $\nu_t^{(c)}$
corresponding to $\Delta=0$, showing that decay of the visibility
signal as a function of the elapsed time is slowed down close to
the critical point. The discontinuity of the derivative of
$\Gamma$ with respect to $\Delta$ can be attributed to the
structural phase transition that the chain undergoes at
$\Delta=0$. In particular, an analytical study shows that for
$\Delta>0$ (on the side of the linear chain) the derivative
$d\Gamma/d\Delta\sim \ln\Delta$ as $\Delta\to 0^+$. This result, whose
derivation is reported in Appendix~\ref{App:B}, is confirmed by a
numerical fit, and is shown in the inset of
Fig.~\ref{fig:shorttimes2}.

As anticipated in Sec.~\ref{sec:discussionspin}, analogous
features have been found in the Loschmidt echo of a spin coupled
to a quantum spin bath, when studying the short-time behaviour of
the echo signal as a function of the control parameter close to
the critical point of the spin chain (see in particular Figs. 2
and 3 of Ref.~\cite{spin-rossini}).

\subsection{Long times: revivals and asymptotic value of the
visibility} \label{Sec:Asymptotics}

\begin{figure} \includegraphics[scale=0.7]{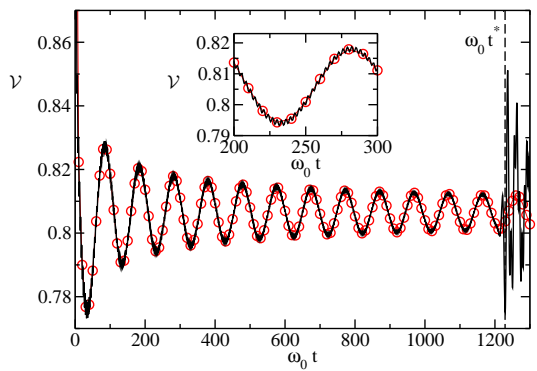}
\caption{\label{fig:vlong} Color online. Behavior of the
visibility signal for long times.  The exact result for the
visibility $\mathcal V$ (solid) obtained using
Eqs.~\eqref{Eq:V} and \eqref{eq:A} and the approximate function
$\exp[-\tilde A_\infty-\tilde B(t)]$ (symbols) (see
Eqs.~\eqref{eq:Aanalytical} and \eqref{eq:Banalytical})  as a function
of time (in units of $1/\omega_0$) for $\Delta=10^{-3}\omega_0$
(linear chain) with $N=10^3$ and $\eta^{(c)}= 0.25$. The vertical
dashed line corresponds to an estimate of the revival time
according to the formula in Eq.~(\ref{t:revival}), which gives
$t^*=1229/\omega_0 $. Inset: the same plot in a smaller region for
a closer comparison between $\mathcal V$ and the approximated
expression $\exp[-\tilde{A}_\infty-\tilde{B}(t)]$. }
 \end{figure}

We finally investigate the behaviour of the visibility ${\mathcal
V}(t)$ when long times $t$ are elapsed between the Ramsey pulses,
focusing on values of the trap frequency $\nu_t$ close to the
critical point. We consider time scales, such that the elapsed
time satisfies the relation  $\omega_{\rm min}t\gg 1$. While the
asynchronous oscillations of the modes of the chain, excited by
the photon recoil, lead to a decay of the visibility as a function
of $t$, in finite systems we expect to observe characteristic
elapsed times $t^*$ where a quasi-synchronization of the motion of
the chain will occur, and, correspondingly, a sort of revival in
the behaviour of the visibility signal is observed. The time $t^*$
can be estimated by considering the propagation speed at which the
mechanical excitation propagates through the crystal and returns
to the initial position. This corresponds to the approximated
formula~\cite{porras-spin-boson}
 \begin{equation} \label{t:revival} t^*=\frac{Na}{v_{max}},
\end{equation} where $v_{max} = \max_k \partial \omega_k /
\partial k$ is the maximum group velocity in the interval $k
\in[0;\pi/a]$. Figure~\ref{fig:vlong} displays the visibility
signal as a function of the elapsed time for a linear chain
composed by $N=10^3$ ions and close to the zigzag instability,
$\Delta=10^{-3}\omega_0$. For $\eta^{(c)}= 0.25$ the group
velocity is maximum at $k=2.64/a$, taking the value $v_{max}=0.81
a \omega_0$. Using this result, the estimated time from
Eq.~\eqref{t:revival} is $t^*=1229/\omega_0$ which approximately
coincides with the time at which the visibility exhibits sudden
oscillations of larger amplitude.

We now focus on the regime, in which the system is sufficiently
large, $N\gg 1$, and derive the asymptotic form of the visibility
signal for times $t\to \infty$, obtained by averaging over time
intervals $T\gg t^*$. The asymptotic form of ${\mathcal V}(t)$ of
can be singled by observing that the exponent $A(t)$ in
Eq.~\eqref{eq:AG} can be rewritten as \begin{equation}
\label{A:long} A=A_\infty+B(t) \end{equation} where
\begin{eqnarray} \label{A:inf}
A_\infty&=&\frac{k_L^2}{2}\left[\langle y_1^2(t)\rangle +\langle
y_1^2(0)\rangle\right]=\sum_{k\sigma}|\alpha_{k\sigma}|^2\,,
\\
B(t)&=&k_L^2 {\rm Re}\langle y_1(t)y_1(0)\rangle=\sum_{k\sigma}
|\alpha_{k\sigma}|^2 \cos\omega_y(k)t\,. \label{B:t}
\end{eqnarray} Using Eq.~(\ref{eq:AG}) we see that $A_\infty$ is
proportional to the variance of the ion displacement $\langle
y_1^2(t)\rangle=\langle y_1^2(0)\rangle$ around its equilibrium
position, while $B(t)$ is proportional to the real part of the
crystal's correlation function $\langle y(t)y(0)\rangle$.

\begin{figure} \includegraphics[scale=0.25]{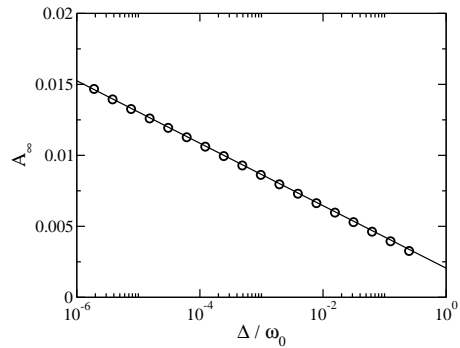}
\caption{\label{fig:ainfty} The asymptotic value of the
correlation function, as given by $A_\infty$ in Eq.~(\ref{A:inf}),
as a function of $\Delta$ when the linear chain is close to the
critical point, where it makes a transition to the zigzag
structure. The symbols correspond to the curve given by
Eq.~\eqref{A:inf} and the solid line is the curve obtained from the
analytical prediction in Eq.~\eqref{eq:Aanalytical}. The
parameters are $N=10^3$ and $\eta^{(c)}\simeq 0.05$.}
 \end{figure}

For $N\gg 1$, and when the linear chain is close to the
instability, $\Delta\to 0^+$, Eqs.~(\ref{A:inf}) and~(\ref{B:t})
can be approximated by the quantities \begin{eqnarray}
\label{eq:Aanalytical} A_\infty&\simeq&\tilde A_\infty=-
\frac{\eta_0^2\nu_t  a}{2\pi h}\ln\Delta + c,
\\
B(t)&\simeq&\tilde B(t)=-\frac{\eta_0^2\nu_t a}{2h}N_0(\delta t),
\label{eq:Banalytical}
\end{eqnarray}
 where $c$ is a
constant, $N_0$ is the Bessel function of the second kind (see
Appendix~\ref{App:B}) and we introduced the parameters
$h=\sqrt{\ln 2}~\omega_0 a$ and \begin{equation}
\delta=\sqrt{\nu_t^2-\nu_t^{(c)2}}, \end{equation} which is
related to the parameter $\Delta$ by the relation
$\delta=\sqrt{\Delta (2\nu_t^{(c)}+\Delta)}$. In
Fig.~\ref{fig:vlong} we compare the exact result of the visibility
with the approximated function $\tilde{\mathcal
V}(t)=\exp[-\tilde{A}_\infty-\tilde{B}(t)]$. For long times, such
that $t\gg 1/\delta $, the function $B(t)$, and as a result the
two-time correlation function of the crystal, decays as
$1/\sqrt{\delta t}$ and the visibility approaches the asymptotic
value given by $V_\infty =\exp[-A_\infty]$.
Figure~\ref{fig:ainfty} displays $A_{\infty}$ as a function of
$\Delta$, showing that $A_\infty$ (and therefore the spatial width
of the ion wavepacket) diverges logarithmically with $\nu_t$ close
to the critical point. As a consequence, the asymptotic value of
the visibility $\mathcal V_\infty$ becomes smaller when the ion
chain is close to the mechanical instability, where it undergoes a
second-order phase transition to the zigzag structure. This
behavior of the visibility is qualitatively similar to the
behavior of the asymptotic Loschmidt echo $\mathcal L_\infty$ for
a spin interacting with an Ising chain close to the critical
magnetic field~\cite{spin-rossini}.

\section{Conclusions} \label{Sec:Conclusions}

The aim of this work was to understand the dynamical properties of a
quasi-one-dimensional many-body system externally driven and
probed via one of its components, in a setup that is both
experimentally feasible and not previously studied in this context
-- a chain of trapped ions. The ions form a Coulomb crystal
undergoing a second-order phase transition between a linear and a
zig-zag configuration, depending on the transverse trapping
strength. Building on interferometric methods with trapped ions
\cite{Sackett}, we proposed a Ramsey experiment to measure the
autocorrelation function of an ion crystal by only exciting and
detecting the internal state of one of its constituent ions.

The central quantity to this study is the visibility of the
fringes of this specific implementation of the Ramsey interferometer.
We show that it provides information of the dynamical and
statistical properties of the crystal. In particular, for the
specific realization here proposed the visibility allows one to
measure the autocorrelation function of the crystal, and the
characteristic decay time, thereby accessing its properties close
to the critical point.

It is interesting to remark that several features of the Ramsey
signal, close to the critical point where the linear chain
undergoes a transition to a zig-zag configuration, share several
analogies with the Loschmidt echo evaluated for studying the
decoherence of a spin coupled to quantum spin-baths close to
criticality~\cite{spin-rossini,paganelli,spin-zanardi,cucchietti}.
In fact, the crystal can be regarded as a reservoir with respect
to the single spin constituted by the two internal states of the
probed ion. In this sense our system can be seen as a model for
decoherence induced by a bath at $T=0$. On the other hand, a
distinguishing feature of our proposal is the presence of
long-range interactions, typically absent in (nearest-neighbor
interacting) spin systems.

We conclude by observing that the ability to drive and probe the
whole crystal by addressing just one ion opens interesting
questions like the possibility of using the methods developed here
to identify control procedures, of spin-echo type, to maintain
coherence in the system, as well as to measure other quantities,
such as, e.g., coherence length and quantum fluctuations at a quantum
critical point. This points to the need for a thorough
understanding of quantum phase transitions in ion traps, as
initiated in \cite{retzker}. The adaptation and further
development of the ideas proposed here to the latter context will
be the subject of our future work.

\acknowledgments We acknowledge stimulating discussions with G.E.
Astrakharchik, Michael Drewsen, J\"urgen Eschner, Michael
Hartmann, Lev P. Pitaevskii, and Efrat Shimshoni. This work has
been supported by the European Commission (EMALI,
MRTN-CT-2006-035369; SCALA, Contract No.\ 015714) and by the
Spanish Ministerio de Educaci\'on y Ciencia (Consolider Ingenio
2010 "QOIT", CSD2006-00019; QLIQS, FIS2005-08257; QNLP,
FIS2007-66944; Ramon-y-Cajal). We wish to acknowledge hospitality
at the Benasque center for science where part of this work has
been developed.

\begin{appendix}

\section{} \label{App:A}

In this appendix we provide more details on the eigenmodes of the
linear and zigzag structures. A more extensive treatment can be
found in~\cite{fishman2008}. For a linear ion chain in a harmonic
trap, the reader is referred to~\cite{morigi2004}.

{\it Linear chain}. For a given frequency $\omega_x(k)$,
$\omega_y(k)$ there are in general two modes of opposite parity
with respect to the transformation $q_j\to q_{-j}$, $y_j\to
y_{-j}$, even $\sigma=+$ and odd $\sigma=-$.
The frequencies at wave vectors $k=0$ and $k=\pi/a$
are instead not degenerate. The corresponding modes are the bulk and the
zigzag mode, respectively, which have definite parity, the first
being even and the latter odd. For the transverse excitations, the
bulk mode is at frequency $\nu_t$ while the zigzag mode is at
frequency $\omega_y(k=\pi/a)$. For the modes at wave vector
$0<k<\pi/a$, the elements of the orthogonal matrix, connecting the
displacements of the ions from the equilibrium positions with the
eigenmodes amplitudes, read \begin{eqnarray}
\label{eq:R} {\cal
R}_{j,k+}&=&\sqrt\frac2N \cos (jka)\,,
\\
{\cal R}_{j,k-}&=&\sqrt\frac2N \sin (jka)\,,
 \end{eqnarray}
while the elements connecting the displacements from the
equilibrium positions with the bulk and zigzag eigenmodes take the
form
 \begin{eqnarray}
{\cal R}_{j,0}&=&\sqrt \frac1N\,,
\\
{\cal R}_{j,\frac{\pi}{a}}&=&(-1)^j\sqrt\frac 1N\,.
 \label{eq:R2}
 \end{eqnarray}

{\it Zigzag configuration.} In the zigzag structure the dispersion
relation is now defined in the Brillouin zone $[0,\pi/2a]$, such
that the wave vector $k$ takes the values $k=2\pi n/Na$ with
$n=0,\ldots,N/4$.
As shown in \cite{fishman2008} the spectrum of excitations consists of four branches that we label $\beta=1,2,3,4$. For a given frequency $\omega_\beta^{zz}(k)$ for $0<k<\pi/2a$ there are two modes with opposite parities $\sigma=\pm$ which are a combination of displacements in the longitudinal and transverse directions.
 The correspondent
elements of the matrix ${\cal R}^{zz}$ read
\begin{eqnarray} {\cal
R}^{zz}_{2j-1,k\beta+} &=&u_\beta \cos (\tilde kaj)\,,
\\
{\cal R}^{zz}_{2j,k\beta+} &=&v_\beta \sin [(\pi-\tilde ka)j]\,,
\\
{\cal R}^{zz}_{2j-1,k\beta-} &=&u_\beta \sin [\tilde kaj]\,,
\\
{\cal R}^{zz}_{2j,k\beta-} &=&-v_\beta \cos [(\pi-\tilde ka)j]\,,
\end{eqnarray}
with $j=1,2,\dots,N$ labeling the ion displacements
$\varrho_{2j-1}=q_j$ and $\varrho_{2j}=w_j$; and $\tilde k =k$ for $\beta=1,2$, $\tilde k = \pi/a-k$
for $\beta=3,4$. The explicit form of the coefficients $u_\beta$
and $v_\beta$ can be found in~\cite{fishman2008} (check Eq.~(20)).

For $k=0$ the normal modes are the bulk modes in the $x$ and $y$ directions. These modes correspond to the zigzag structure oscillating rigidly in the $x$ or $y$ directions. The matrix elements for the bulk mode in the $x$ direction are:
\begin{equation}
{\cal R}^{zz}_{2j-1,0x}=\sqrt \frac1N; \quad {\cal R}^{zz}_{2j,0x}=0;
\end{equation}
while for the bulk mode in the $y$ direction they are
\begin{equation}
{\cal R}^{zz}_{2j-1,0y}=0; \quad {\cal R}^{zz}_{2j,0y}=\sqrt \frac1N.
\end{equation}

For $k=\pi/2a$ the normal modes are the two zigzag modes along the $x$ and $y$ direction. The zigzag mode in the $x$ direction is the mode where  neighboring ions oscillate around the equilibrium positions along the $x$ axis and with opposite phase. The corresponding matrix elements are given by:
\begin{equation}
{\cal R}^{zz}_{2j-1,\frac{\pi}{2a}\,x}=(-1)^j\sqrt\frac 1N;\quad {\cal R}^{zz}_{2j,\frac{\pi}{2a}\,x}=0.
\end{equation}
Analogously, for the zigzag mode in the $y$ direction, the matrix elements are:
\begin{equation}
{\cal R}^{zz}_{2j-1,\frac{\pi}{2a}\,y}=0;\quad {\cal R}^{zz}_{2j,\frac{\pi}{2a}\,y}=(-1)^j\sqrt\frac 1N.
\end{equation}

\section{}
\label{App:B}

In this section we determine analytically the behaviour of the
function $A(t)$, Eq.~(\ref{eq:A}), for short and long elapsed
times $t$. We focus on the linear chain, when the parameters are
such that $\nu_t$ is close to the value $\nu_t^{(c)}$ of the
transition to the zigzag configuration.

{\it Short time behaviour.} For short times, such that $\nu_tt\ll
1$, we expand $A(t)$ at lowest order in the small parameter
$\omega_y(k)t\ll 1$, obtaining Eq.~(\ref{eq:Agauss}), where
$\Gamma$ is given by expression
\eqref{eq:expansion-simple-ramsey}. In the linear chain ($\Delta>0$), using
the definition of $\alpha_{k\sigma}$ in Eq.~\eqref{eq:alphaksigma} with the
corresponding values of the elements of the matrix ${\cal R}$ in
Eqs.~\eqref{eq:R}-\eqref{eq:R2}, we find
\begin{eqnarray}
\Gamma&=&\frac{\hbar k_L^2}{4m}\sum_{k,\sigma} \mathcal R_{1, k\sigma}^2 \omega_y(k)=
\\
&=& \frac{\hbar k_L^2}{4m}\left(
\frac{1}{N}\sum_{k,\sigma}\omega_y(k)\right)\,.
\label{eq:gamma1}
\end{eqnarray}
The value of $\Gamma$ is hence proportional to the mean value of
the transverse frequencies.

An explicit form of the mean value can be obtained close to the
phase transition, when $\Delta=\nu_t-\nu_t^{(c)}\ll\nu_t^{(c)}$,
and for a sufficiently large number of ions. In this limit, we can
approximate the dispersion relation Eq.~\eqref{eq:linspectrum2}
for small momenta $q=\pi/a-k$ as: \begin{equation} \label{eq:wq}
\omega_y(q) \simeq \sqrt{\delta^2+h^2q^2}\,, \end{equation} where
$\delta=\sqrt{\nu_t^2-\nu_t^{(c)2}}\simeq
\sqrt{2\nu_t^{(c)}\Delta}$ and $h=\sqrt{\ln 2}\omega_0 a$. Taking
the continuum limit in Eq.~\eqref{eq:gamma1} we find
\begin{eqnarray}
\Gamma&\simeq& \frac 12 \eta_0^2\nu_t\frac
a\pi\int_0^{\pi/a} dk
\omega_y(k)\nonumber\\
&\simeq& \frac 12 \eta^2_0 \nu_t \frac a\pi \int_0^{q_{\rm max}}
\sqrt{\delta^2 +h^2 q^2} dq+c
\nonumber\\
&=&\frac 14 \eta^2_0 \nu_t \frac a\pi \frac 1h \left[h q_{\rm
max}\sqrt {\delta^2+h^2 q_{\rm max}^2}+
 \right .\nonumber\\
&+&  \left .
 \delta^2 \textrm{arcsinh}\left( \frac{h q_{\rm max}}{\delta} \right)  \right] +
 c\,, \label{Gamma:Delta}
\end{eqnarray} where $q_{\rm max}$ is a cutoff and $c$ is a
constant which depends on the cutoff. The derivative of $\Gamma$
with respect to $\Delta$, for $\Delta\to 0^+$ is obtained from
Eq.~(\ref{Gamma:Delta}) and reads \begin{equation}
\frac{d\Gamma}{d\Delta}=-\frac 14\eta^2_0 \nu_t \frac{\nu_t^{(c)}
a }{\pi h}\ln\Delta +c_1\,, \end{equation} where $c_1$ depends on
$q_{\rm max}$ and and we neglected terms which goes to zero as
$\Delta\to 0$.

{\it Long time behaviour.}  We now study the behaviour of the
function $A(t)$ for long times. The quantity $A_\infty$ can be
cast in a simplified form for the linear chain. Using the same technique which leads to Eq.~\eqref{eq:gamma1}, one finds
 \begin{equation}
A_\infty=\frac{\hbar k_L^2}{2m}\left(\frac{1}{N}\sum_{k,\sigma}\frac{1}{\omega_y(k)}\right)\,.
\end{equation} The value close to the critical point and for large
chains can be found following the same procedure applied for
obtaining Eq.~(\ref{Gamma:Delta}). We get \begin{eqnarray}
A_\infty&\simeq& \frac{\hbar k_L^2}{2m}\frac a\pi \int_0^{q_{\rm
max}}
\frac{dq}{\sqrt{\delta^2+h^2 q^2}}\nonumber\\
&\simeq&- \frac{\eta_0^2\nu_t  a}{2\pi h}\ln\Delta + c\,,
\end{eqnarray} where $c$ is a constant which depends on $q_{\rm
max}$ and we omitted terms which vanish when $\delta\to 0$. We now
turn to the expression \eqref{B:t} for $B(t)$ and with the aid of
formula \eqref{eq:wq} we can approximate it as: \begin{eqnarray}
B(t)&=&\frac{\eta_0^2\nu_t a}{\pi}\int_0^{\infty}
\frac{\cos\omega_y(q)t}{\omega_y(q)}dq
\\
&=& -\frac{\eta_0^2\nu_t a}{2h}N_0(\delta t)\,, \end{eqnarray}
where we let $q_{\rm max}\to\infty$. The function $N_0(x)$ is the
Bessel function of the second kind (also von Neumann function)
\cite{grad}. At the asymptotics, for $x\to\infty$, it behaves as
\cite{grad} \begin{equation} N_0(x) \simeq \frac{\sin x-\cos
x}{\sqrt{\pi x}}\,, \end{equation} and hence vanishes as
$1/\sqrt{x}$. \end{appendix}

 \end{document}